\documentclass{article}

\usepackage{arxiv}

\usepackage[utf8]{inputenc} 
\usepackage[T1]{fontenc}    
\usepackage{fancyhdr}
\usepackage{lmodern}
\usepackage[colorlinks]{hyperref}  
\usepackage{url}            
\usepackage{booktabs}       
\usepackage{amsfonts}       
\usepackage{nicefrac}       
\usepackage{microtype}      
\usepackage{lipsum}

\usepackage{graphicx}
\usepackage{epsfig}
\usepackage{epstopdf}
\usepackage{subfigure}

\usepackage{enumerate}
\usepackage{amssymb}
\usepackage{amsmath}
\usepackage{amsthm}
\usepackage{enumitem}
\usepackage{mathrsfs}

\newcommand{\R}{\mathbb{R}}

\newcommand{\N}{\mathbb{N}}
\newcommand{\T}{\mathcal{T}}

\renewcommand{\bar}{\overline}
\renewcommand{\rho}{\varrho}
\renewcommand{\phi}{\varphi}

\title{The effects of Vadasz term, anisotropy and rotation on bi-disperse convection}

\author{  
F. Capone\thanks{Corresponding author.} \href{https://orcid.org/0000-0002-0672-999X}{\includegraphics[scale=0.1]{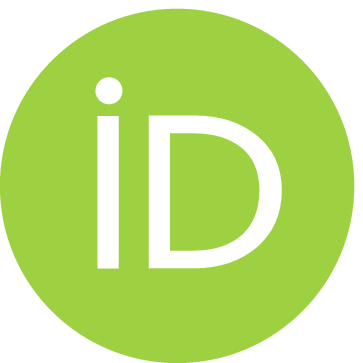}} \\ Dipartimento di Matematica e Applicazioni 'R.Caccioppoli' \\ Università degli Studi di Napoli Federico II \\ Via Cintia, Monte S.Angelo, 80126 Napoli \\ Italy \\ 
\texttt{fcapone@unina.it} \\
\And  
G. Massa\href{https://orcid.org/0000-0002-8401-9176}{\includegraphics[scale=0.1]{orcid.eps}} \\ Dipartimento di Matematica e Applicazioni 'R.Caccioppoli' \\ Università degli Studi di Napoli Federico II \\ Via Cintia, Monte S.Angelo, 80126 Napoli \\ Italy \\   
\texttt{giuliana.massa@unina.it} \\ }

\begin{document}
\maketitle

\begin{abstract}
The onset of thermal convection in a uniformly rotating and horizontally isotropic bi-disperse porous medium, taking into account the Vadasz term, is investigated. Via linear instability analysis, it has been proven that the Vadasz term allows the onset of convection via an oscillatory state but does not affect convection via a stationary motion.
\end{abstract}

\keywords{Anisotropy \and Bi-disperse Porous Media \and Rotating Layer \and Vadasz Term \and Instability analysis}

\section{Introduction} 
A \textit{bi-disperse porous medium} (BDPM) is a dual porosity material characterized by a standard pore structure, but the solid skeleton has fractures or cracks in it. In particular, a BDPM is a compound of {\it clusters} of large particles that are themselves aggregations of smaller particles: the macropores between the clusters are referred to as \textit{f-phase} (meaning fracture phase) and have porosity $\phi$, while the remainder of the structure is referred to as \textit{p-phase} (meaning porous phase) and the porosity of the micropores within the clusters is $\epsilon$ \cite{NK2006}. \\ \\
While thermal convection has been widely studied in both clear fluid and porous media by many authors (see for example \cite{cap-gent, caponeHill, chandr, tyvand, NB, Vadasz, Vadasz2} and references therein), in the past as nowadays, double porosity materials has recently attracted many researchers due to their applications in engineering, medical field, chemistry (see \cite{BSGentile, BSGentile2, brianLibro} and references therein) and a theoretical key development is  attributable to Nield and Kuznetsov in \cite{NK2005-1,NK2005,NK2006}. Bi-disperse porous media may be pretty useful in a laboratory \cite{NK2006}, but in particular \textit{anisotropic} bi-disperse porous materials offer much more possibilities to design man-made materials for heat transfer or insulation problems \cite{legg, brian1, brian3}. \\ The analysis of fluid motion in rotating porous media finds many applications in geophysics and in engineering, for example for rotating machinery, chemical process industry, centrifugal filtration processes (see \cite{Vadasz} and references therein), hence the study of thermal convection in rotating BDPM may be necessary and useful as well (see \cite{coriolis, legg, inertia}). \\ Regarding the \textit{Vadasz term} effect, it has been largely analysed by many authors in single porosity media (see for instance \cite{MHD, doublediff, cap-gent, tyvand, Vadasz2}) since the Vadasz number has a remarkable effect on the onset of convection in a rotating porous layer, in particular in \cite{Vadasz2} it has been proved that the Vadasz term leads to the onset of convection via an oscillatory state. On the other hand, the effect of Vadasz number on the onset of bi-disperse convection has been investigated by Straughan in \cite{BSinertia}, where he considered a fluid mixture saturating a BDPM, and by Capone and De Luca in \cite{inertia}, which deals with an isotropic and rotating BDPM. \\ \\ 
The goal of the present paper is to analyse the combined effects of anisotropic permeabilities, uniform rotation about a vertical axis and inertia on the onset of thermal convection in an incompressible fluid saturating a single temperature bi-disperse porous medium. The paper is organized as follows. In section \ref{model} the mathematical model and the associated perturbation equations are introduced. In section \ref{linear} we perform linear instability analysis of the thermal conduction solution, in particular, we find out that the Vadasz term allows the onset of thermal convection via an oscillatory state (named as oscillatory convection), but it does not affect the onset of thermal convection via a steady state (named as stationary convection). In sections \ref{steady} and \ref{oscill} we determine the critical Rayleigh numbers for the onset of steady and oscillatory convection, respectively. In section \ref{numerical} we perform numerical simulations in order to analyse the behaviour of the instability thresholds with respect to fundamental parameters. The paper ends with a concluding section that recaps all the results.   

\section{Mathematical model}\label{model}
Let $Oxyz$ be a reference frame with fundamental unit vectors ${\bf i,j,k}$ and let us assume that the plane layer $L$, of thickness $d$, of saturated bi-disperse porous medium is uniformly heated from below and rotates about the vertical axis $z$, let ${\bf \Omega}=\Omega {\bf k}$ be the constant angular velocity of the layer. Furthermore, we consider a single temperature bi-disperse porous medium, i.e. $T^f=T^p=T$. We restrict our attention to the case in which the permeabilities of the saturated bi-disperse porous medium are \textit{horizontally isotropic}.\\ 
Let the axes $(x,y,z)$ be the \textit{principal axes} of the permeabilities, so the macropermeability tensor and the micropermeability tensor may be written as

$$ {\bf K}^f = \text{diag}(K^f_x,K^f_y,K^f_z)=K^f_z \ {\bf K}^{f*},$$
$$ {\bf K}^p = \text{diag}(K^p_x,K^p_y,K^p_z)=K^p_z \ {\bf K}^{p*},$$
$${\bf K}^{f*}=\text{diag}(k,k,1),$$
$${\bf K}^{p*}=\text{diag}(h,h,1),$$
where $$k=\frac{K^f_x}{K^f_z}=\frac{K^f_y}{K^f_z},$$
$$h=\frac{K^p_x}{K^p_z}=\frac{K^p_y}{K^p_z}.$$

Darcy's model, with the Oberbeque-Boussinesq approximation, is employed, in particular it is extended in both micropores and macropores in order to include the Coriolis terms and is extended only in the macropores to include the time derivative term of the seepage velocity (see \cite{inertia, BSinertia}). The equations describing the evolutionary behaviour of thermal convection in a rotating horizontally isotropic bi-disperse porous medium are, cf. \cite{coriolis, inertia},
\begin{equation} \label{sist1}
\begin{cases} \rho_F c_a \dfrac{\partial {\bf v}^f}{\partial t} = - \mu ({\bf K}^f)^{-1} {\bf v}^f - \zeta ({\bf v}^f - {\bf v}^p) - \nabla p^f + \rho_F \alpha g T {\bf k} - \dfrac{2 \rho_F \Omega}{\phi} {\bf k} \times {\bf v}^f, \\  - \mu ({\bf K}^p)^{-1} {\bf v}^p - \zeta ({\bf v}^p - {\bf v}^f) - \nabla p^p + \rho_F \alpha g T {\bf k} - \dfrac{2 \rho_F \Omega}{\epsilon} {\bf k} \times {\bf v}^p = {\bf 0},  \\  \nabla \cdot {\bf v}^f = 0, \\  \nabla \cdot {\bf v}^p = 0, \\  (\rho c)_m \dfrac{\partial T}{\partial t} + (\rho c)_f ({\bf v}^f + {\bf v}^p) \cdot \nabla T = k_m \Delta T, \end{cases}
\end{equation}
where 
$$p^s=P^s-\frac{\rho_F}{2} \vert {\bf \Omega} \times {\bf x} \vert^2, \quad s=\{f,p\}$$ are the reduced pressures, ${\bf x}=(x,y,z)$, ${\bf v}^s$ = seepage velocity for $s=\{f,p\}$, $\zeta$ = interaction coefficient between the f-phase and the p-phase, ${\bf g}=-g {\bf k}$ = gravity, $\mu$ = fluid viscosity, $\rho_F$ = reference constant density, $\alpha$ = thermal expansion coefficient, $c$ = specific heat, $c_p$ = specific heat at a constant pressure, $c_a$= acceleration coefficient, $(\rho c)_m=(1-\phi)(1-\epsilon)(\rho c)_{sol}+\phi (\rho c)_f+\epsilon (1-\phi)(\rho c)_p$, $k_m=(1-\phi)(1-\epsilon)k_{sol}+\phi k_f+\epsilon (1-\phi)k_p$ = thermal conductivity (the subscript $sol$ is referred to the solid skeleton).  

To $\ref{sist1}$ the following boundary conditions are appended
\begin{equation} \label{BC1} \begin{array}{l} {\bf v}^s \cdot {\bf n} = 0 \ \text{on} \ z=0,d, \ \text{for} \ s=\{f,p\} \\ T=T_L \ \text{on} \ z=0, \quad T=T_U \ \text{on} \ z=d, \end{array} \end{equation} 
where ${\bf n}$ is the unit outward normal to the impermeable horizontal planes delimiting the layer and $T_L>T_U$. \\ \\ 
System $(\ref{sist1})$-$(\ref{BC1})$ admits the stationary conduction solution:
$${\bf \overline{v}}^f=0, \ {\bf \overline{v}}^p=0, \ \overline{T}=- \beta z + T_L,$$
where $\beta=\dfrac{T_L-T_U}{d}$ is the temperature gradient.
Denoting by $\{ {\bf u}^f, {\bf u}^p, \theta, \pi^f, \pi^p \}$ a generic perturbation to the steady solution, the resulting perturbation equations are
\begin{equation}
\begin{cases}  \rho_F c_a \dfrac{\partial {\bf u}^f}{\partial t} = - \mu ({\bf K}^f)^{-1} {\bf u}^f - \zeta ({\bf u}^f - {\bf u}^p) - \nabla \pi^f + \rho_F \alpha g \theta {\bf k} - \dfrac{2 \rho_F \Omega}{\phi} {\bf k} \times {\bf u}^f, \\  - \mu ({\bf K}^p)^{-1} {\bf u}^p - \zeta ({\bf u}^p - {\bf u}^f) - \nabla \pi^p + \rho_F \alpha g \theta {\bf k} - \dfrac{2 \rho_F \Omega}{\epsilon} {\bf k} \times {\bf u}^p = {\bf 0},  \\  \nabla \cdot {\bf u}^f = 0, \\  \nabla \cdot {\bf u}^p = 0, \\  (\rho c)_m \dfrac{\partial \theta}{\partial t} + (\rho c)_f ({\bf u}^f + {\bf u}^p) \cdot \nabla \theta = (\rho c)_f \beta (w^f+w^p) + k_m \Delta \theta.  \end{cases} \label{pertsist}
\end{equation}
where ${\bf u}^f=(u^f,v^f,w^f)$ and ${\bf u}^p=(u^p,v^p,w^p)$. The above system $(\ref{pertsist})$ may be non-dimensionalized with the following non-dimensional parameters
$$ \begin{array}{l} 
{\bf x}^{*} = \dfrac{{\bf x}}{d}, \ t^{*}=\dfrac{t}{\tilde{t}}, \ \theta^{*} = \dfrac{\theta}{\tilde{T}}, \\ {\bf u}^{s*} = \dfrac{{\bf u}^s}{\tilde{u}}, \ \pi^{s*} = \dfrac{\pi^s}{\tilde{P}}, \ \text{for} \ s=\{ f,p \}
\\ \eta=\dfrac{\phi}{\epsilon}, \ \gamma=\dfrac{K^f_z \zeta}{\mu}, \ K_r=\dfrac{K^f_z}{K^p_z}, \end{array}  $$

where the scales are given by 
$$\tilde{u}=\frac{k_m}{(\rho c)_f d}, \ \tilde{t} = \frac{d^2 (\rho c)_m}{k_m}, \ \tilde{P} = \frac{\mu k_m}{(\rho c)_f K_z^f}, \ \tilde{T} = \sqrt{\frac{\beta k_m \mu}{(\rho c)_f \rho_F \alpha g K_z^f}}.$$
The resulting non-dimensional perturbation equations, dropping all the asterisks, are 
\begin{equation}
\begin{cases}  - J \dfrac{\partial {\bf u}^f}{\partial t} -({\bf K}^f)^{-1} {\bf u}^f - \gamma ({\bf u}^f - {\bf u}^p) - \nabla \pi^f + R \theta {\bf k} - \mathcal{T} {\bf k} \times {\bf u}^f = {\bf 0}, \\  -K_r ({\bf K}^p)^{-1} {\bf u}^p - \gamma ({\bf u}^p - {\bf u}^f) - \nabla \pi^p + R \theta {\bf k} - \eta \mathcal{T} {\bf k} \times {\bf u}^p = {\bf 0}, \\  \nabla \cdot {\bf u}^f = 0, \\  \nabla \cdot {\bf u}^p = 0, \\  \dfrac{\partial \theta}{\partial t} + ({\bf u}^f + {\bf u}^p) \cdot \nabla \theta = w^f+w^p +\Delta \theta, \end{cases} \label{pertubations}
\end{equation}
where the Taylor number $\T$, the Vadasz number $J$ and the Rayleigh number $R$ are
$$\T = \dfrac{2 \rho_F \Omega K^f_z}{\phi \mu}, \qquad J=\dfrac{K^f_z \rho_F c_a k_m}{\mu d^2 (\rho c)_m}, \qquad R = \sqrt{\dfrac{\beta d^2 (\rho c)_f \rho_F \alpha g K_z^f}{k_m \mu}}.$$
To system $(\ref{pertubations})$ the following initial and boundary conditions are appended: 
$${\bf u}^s({\bf x},0)={\bf u}^s_0({\bf x}), \quad \pi^s({\bf x},0)=\pi^s_0({\bf x}), \quad \theta({\bf x},0)=\theta_0({\bf x}),$$
with $\nabla \cdot {\bf u}^s_0=0$, for $s=\{f,p\}$, and
\begin{equation} \label{BC}
w^f=w^p=\theta=0 \quad \text{on} \ z=0,1.
\end{equation}
According to experimental results, the solutions are required to be periodic in the horizontal directions $x$ and $y$ and in the sequel we will denote by $$V=\Big[ 0,\frac{2 \pi}{l} \Big] \times \Big[ 0,\frac{2 \pi}{m} \Big] \times [0,1]$$ the periodicity cell.

\section{Onset of convection}\label{linear}
In order to determine the linear instability threshold of the thermal conduction solution, let us linearise system $(\ref{pertubations})$, i.e. 
\begin{equation}
\begin{cases}  - J \dfrac{\partial {\bf u}^f}{\partial t} -({\bf K}^f)^{-1} {\bf u}^f - \gamma ({\bf u}^f - {\bf u}^p) - \nabla \pi^f + R \theta {\bf k} - \mathcal{T} {\bf k} \times {\bf u}^f = {\bf 0}, \\  -K_r ({\bf K}^p)^{-1} {\bf u}^p - \gamma ({\bf u}^p - {\bf u}^f) - \nabla \pi^p + R \theta {\bf k} - \eta \mathcal{T} {\bf k} \times {\bf u}^p = {\bf 0}, \\  \nabla \cdot {\bf u}^f = 0, \\  \nabla \cdot {\bf u}^p = 0, \\  \dfrac{\partial \theta}{\partial t} = w^f+w^p +\Delta \theta. \end{cases} \label{pertLIN}
\end{equation}
Being the system $(\ref{pertLIN})$ autonomous, we seek solutions with time-dependence like $e^{\sigma t}$, i.e. 
\begin{equation} \label{sol}
\begin{aligned} {\bf u}^s(t,{\bf x}) &= e^{\sigma t} {\bf u}^s({\bf x}), \\ \theta(t,{\bf x}) &= e^{\sigma t} \theta({\bf x}), \\ \pi^s(t,{\bf x}) &= e^{\sigma t} \pi^s({\bf x}), \end{aligned}
\end{equation}
with $\sigma \in \mathbb{C}$ and $s=\{f,p\}$. By virtue of $(\ref{sol})$, $(\ref{pertLIN})$ becomes
\begin{equation}
\begin{cases}  - J \sigma {\bf u}^f -({\bf K}^f)^{-1} {\bf u}^f - \gamma ({\bf u}^f - {\bf u}^p) - \nabla \pi^f + R \theta {\bf k} - \mathcal{T} {\bf k} \times {\bf u}^f = {\bf 0}, \\  -K_r ({\bf K}^p)^{-1} {\bf u}^p - \gamma ({\bf u}^p - {\bf u}^f) - \nabla \pi^p + R \theta {\bf k} - \eta \mathcal{T} {\bf k} \times {\bf u}^p = {\bf 0}, \\  \nabla \cdot {\bf u}^f = 0, \\  \nabla \cdot {\bf u}^p = 0, \\  \sigma \theta = w^f+w^p +\Delta \theta. \end{cases} \label{pert}
\end{equation}
Setting
$$\omega^s_3=(\nabla \times {\bf u}^s) \cdot {\bf k},\,\, s=\{f,p\},$$
let us consider the third components of the curl and of the double curl of $(\ref{pert})_{1,2}$, i.e.
\begin{equation}\label{curl} \begin{cases}
(J \sigma + \dfrac{1}{k} + \gamma) \omega_3^f - \gamma \omega^p_3 -\T \dfrac{\partial w^f}{\partial z} =0, \\ (K_r \dfrac{1}{h} + \gamma) \omega_3^p - \gamma \omega^f_3 - \eta \T \dfrac{\partial w^p}{\partial z} =0,  \end{cases}
\end{equation}
and
\begin{equation}\label{doublecurl} \begin{cases}
(J \sigma + \dfrac{1}{k} + \gamma) \dfrac{\partial^2 w^f}{\partial z^2} +(J \sigma+1+\gamma) \Delta_1 w^f - \gamma \Bigl( \Delta_1 w^p + \dfrac{\partial^2 w^p}{\partial z^2} \Bigr) - R \Delta_1 \theta + \T \dfrac{\partial \omega_3^f}{\partial z} = 0, \\ (K_r \dfrac{1}{h} + \gamma) \dfrac{\partial^2 w^p}{\partial z^2} + (K_r+\gamma) \Delta_1 w^p - \gamma \Bigl( \Delta_1 w^f + \dfrac{\partial^2 w^f}{\partial z^2} \Bigr) -R \Delta_1 \theta + \eta \T \dfrac{\partial \omega_3^p}{\partial z} =0,  \end{cases}
\end{equation}
where $\Delta_1 \equiv \partial^2 / \partial x^2 + \partial^2 / \partial y^2$ is the horizontal laplacian. Solving $(\ref{curl})$ with respect to $\omega_3^f$ and $\omega_3^p$, one obtains
\begin{equation} \label{curl1}
\begin{cases} \omega_3^f= \dfrac{B_1}{H} \T \dfrac{\partial w^f}{\partial z} + \dfrac{\gamma}{H} \eta \T \dfrac{\partial w^p}{\partial z}, \\ \omega_3^p= \dfrac{\gamma}{H} \T \dfrac{\partial w^f}{\partial z} + \dfrac{A_1}{H} \eta \T \dfrac{\partial w^p}{\partial z}, \end{cases}
\end{equation}
where
$$\begin{aligned} A_1 &= J \sigma + \dfrac{1}{k} + \gamma, \\ B_1 &= K_r \dfrac{1}{h} +\gamma, \\ H &= \Bigl( J \sigma +\dfrac{1}{k} \Bigr) \Bigl( K_r \dfrac{1}{h} + \gamma \Bigr) + K_r \dfrac{1}{h} \gamma. \end{aligned} $$
Substituting the derivative with respect to $z$ of $(\ref{curl1})$ into $(\ref{doublecurl})$, one gets
\begin{equation}\label{doublecurl1} \begin{cases}
\Bigl( A_1+ \dfrac{B_1}{H} \T^2 \Bigr) \dfrac{\partial^2 w^f}{\partial z^2} +A_2 \Delta_1 w^f - \gamma \Delta_1 w^p + \Bigl( \dfrac{\gamma}{H} \eta \T^2 -\gamma \Bigr) \dfrac{\partial^2 w^p}{\partial z^2} - R \Delta_1 \theta = 0, \\ \Bigl( \dfrac{\gamma}{H} \eta \T^2 -\gamma \Bigr) \dfrac{\partial^2 w^f}{\partial z^2} -\gamma \Delta_1 w^f +B_2 \Delta_1 w^p + \Bigl(B_1+ \dfrac{A_1}{H} \eta^2 \T^2 \Bigr) \dfrac{\partial^2 w^p}{\partial z^2} - R \Delta_1 \theta = 0,  \end{cases}
\end{equation}
with $A_2=J \sigma + 1 +\gamma, \ B_2= K_r+\gamma$.
Hence, considering $(\ref{doublecurl1})_{1,2}$ and $(\ref{pert})_5$ we get the following problem in $w^f,w^p,\theta$
\begin{equation}\label{probl} \begin{cases}
\Bigl( A_1+ \dfrac{B_1}{H} \T^2 \Bigr) \dfrac{\partial^2 w^f}{\partial z^2} +A_2 \Delta_1 w^f - \gamma \Delta_1 w^p + \Bigl( \dfrac{\gamma}{H} \eta \T^2 -\gamma \Bigr) \dfrac{\partial^2 w^p}{\partial z^2} - R \Delta_1 \theta = 0, \\ \Bigl( \dfrac{\gamma}{H} \eta \T^2 -\gamma \Bigr) \dfrac{\partial^2 w^f}{\partial z^2} -\gamma \Delta_1 w^f +B_2 \Delta_1 w^p + \Bigl(B_1+ \dfrac{A_1}{H} \eta^2 \T^2 \Bigr) \dfrac{\partial^2 w^p}{\partial z^2} - R \Delta_1 \theta = 0, \\ \sigma \theta = w^f+w^p +\Delta \theta.  \end{cases}
\end{equation}
According to the boundary conditions $(\ref{BC})$ and to the periodicity of the perturbations fields, being $\{ \sin(n \pi z) \}_{n \in \N}$ a complete orthogonal system for $L^2([0,1])$, we seek for normal modes solutions
\begin{equation}
\begin{aligned} w^f &= W^f_0 \sin(n \pi z) e^{i(lx+my)}, \\ w^p &= W^p_0 \sin(n \pi z) e^{i(lx+my)}, \\ \theta &= \Theta_0 \sin(n \pi z) e^{i(lx+my)},
\end{aligned}
\end{equation}
with $W^f_0,W^p_0,\Theta_0$ real constants. Hence, employing normal modes solutions, system $(\ref{probl})$ becomes
\begin{equation}
\begin{cases}
- \Bigl[ \Bigl( A_1+\dfrac{B_1}{H} \T^2 \Bigr) n^2 \pi^2 +A_2 a^2 \Bigr] W^f_0 + \Bigl[ \gamma \Lambda_n - \dfrac{\gamma}{H} \eta \T^2 n^2 \pi^2 \Bigr] W^p_0 + R a^2 \Theta_0=0, \\ \Bigl[ \gamma \Lambda_n - \dfrac{\gamma}{H} \eta \T^2 n^2 \pi^2 \Bigr] W^f_0 - \Bigl[ \Bigl( B_1+\dfrac{A_1}{H} \eta^2 \T^2 \Bigr) n^2 \pi^2 +B_2 a^2 \Bigr] W^p_0 + R a^2 \Theta_0=0, \\ W^f_0+W^p_0-\Theta_0(\Lambda_n+\sigma)=0, \end{cases}
\end{equation}
i.e. 
\begin{equation}\label{det}
\begin{cases} - \Bigl[ (J \sigma+\gamma) \Lambda_n + \Lambda_n^k +\dfrac{B_1}{H} \T^2 n^2 \pi^2 \Bigr] W^f_0 +  \Bigl[ \gamma \Lambda_n - \dfrac{\gamma}{H} \eta \T^2 n^2 \pi^2 \Bigr] W^p_0 +R a^2 \Theta_0=0 \\  \Bigl[ \gamma \Lambda_n - \dfrac{\gamma}{H} \eta \T^2 n^2 \pi^2 \Bigr] W^f_0 - \Bigl[ \gamma \Lambda_n + K_r \Lambda_n^h + \dfrac{A_1}{H} \eta^2 \T^2 n^2 \pi^2 \Bigr] W^p_0 + R a^2 \Theta_0=0, \\ W^f_0+W^p_0-\Theta_0(\Lambda_n+\sigma)=0, \end{cases}
\end{equation}
where $a^2=l^2+m^2$ is the wavenumber and
$$\Lambda_n=n^2 \pi^2 + a^2, \quad \Lambda_n^k=\dfrac{1}{k} n^2 \pi^2 + a^2, \quad \Lambda_n^h = \dfrac{1}{h} n^2 \pi^2 + a^2.$$
Requiring zero determinant for $(\ref{det})$, one obtains
\begin{equation}\label{soglia}
\! R \!=\! \dfrac{\Lambda_n \!+\! \sigma}{a^2} \dfrac{H J \gamma \sigma \Lambda_n^2 \!+\! \Bigl[\gamma \Lambda_n^k \!+\! (J \sigma \!+\! \gamma) K_r \Lambda_n^h \Bigr] H \Lambda_n \!+\! K_r H \Lambda_n^k \Lambda_n^h \!+\! M \Lambda_n \T^2 n^2 \pi^2 \!+\! \eta^2 \T^4 n^4 \pi^4 \!+\! N \T^2 n^2 \pi^2}{H(J \sigma+4\gamma)\Lambda_n+H(\Lambda_n^k+K_r\Lambda_n^h)+E\T^2 n^2 \pi^2}
\end{equation}
where the following positions have been made
$$\begin{aligned} M &= \gamma^2 (\eta+1)^2+ \eta^2 \Bigl( J \sigma+2 \gamma +\dfrac{1}{k} \Bigr) J \sigma + \gamma \Bigl( \dfrac{1}{k} \eta^2 +K_r \dfrac{1}{h} \Bigr), \\ N &= \eta^2 \Bigl( J \sigma + \dfrac{1}{k} + \gamma \Bigr) \Lambda_n^k + K_r \Bigl( K_r \dfrac{1}{h} +\gamma \Bigr) \Lambda_n^h, \\ E &= \gamma (\eta-1)^2 + K_r \dfrac{1}{h} +\eta^2 \Bigl( J \sigma+\dfrac{1}{k} \Bigr). 
\end{aligned}$$

\subsection{Steady convection} \label{steady}
In order to determine the instability threshold for the onset of steady convection, let us set $\sigma=0$ into $(\ref{soglia})$. Then the critical Rayleigh number for the onset of stationary convection is given by 
\begin{equation} \label{staz}
R_S \!=\! \min_{(n,a^2) \in \N \times \R^+} \dfrac{\Lambda_n}{a^2} \dfrac{\gamma ( \Lambda_n^k \!+\! K_r \Lambda_n^h ) Q \Lambda_n \!+\! K_r Q \Lambda_n^k \Lambda_n^h \!+\! M_S \Lambda_n \T^2 n^2 \pi^2 \!+\! \eta^2 \T^4 n^4 \pi^4 \!+\! N_S \T^2 n^2 \pi^2}{Q 4 \gamma \Lambda_n+Q (\Lambda_n^k+K_r\Lambda_n^h)+E_S \T^2 n^2 \pi^2} \ ,
\end{equation}
where
$$\begin{aligned} Q &=\gamma \Bigl( K_r \dfrac{1}{h} + \dfrac{1}{k} \Bigr) +K_r \dfrac{1}{hk}, \\ M_S &= \gamma^2 (\eta+1)^2 + \gamma \Bigl( \dfrac{1}{k} \eta^2 +K_r \dfrac{1}{h} \Bigr), \\ N_S &= \eta^2 \Bigl( \dfrac{1}{k} + \gamma \Bigr) \Lambda_n^k + K_r \Bigl( K_r \dfrac{1}{h} +\gamma \Bigr) \Lambda_n^h, \\ E_S &= \gamma (\eta-1)^2 + K_r \dfrac{1}{h} +\eta^2 \dfrac{1}{k}, 
\end{aligned}$$
the solution of $(\ref{staz})$ will be analysed through numerical simulations in Section \ref{numerical}.
Let us observe that
\begin{itemize}
\item[i)] $R_S$ does not depend on $J$, and hence the Vadasz term does not affect  the onset of steady convection; 
\item[ii)] Since $\dfrac{\partial R_S}{\partial \T^2}>0$, as expected, rotation delays the onset of stationary convection. 
\end{itemize}

\subsection{Oscillatory convection} \label{oscill}
In order to determine the oscillatory convection threshold, setting $\sigma=i \sigma_1$, with $\sigma_1 \in \R-\{0\}$, and
$$ \Psi = \Lambda_n^k + K_r \Lambda_n^h, \quad  \Gamma = K_r \dfrac{1}{h} +\gamma, $$
$(\ref{soglia})$ becomes
\begin{equation}\label{soglia2}
R=\dfrac{\Lambda_n+i \sigma_1}{a^2} \dfrac{f}{g},
\end{equation}
with 
$$\begin{aligned}
f =& \Lambda_n \Bigl( -\Lambda_n^h J^2 \sigma_1^2 \Gamma K_r + \gamma \Psi Q \Bigr) - \Gamma J^2 \sigma_1^2 \gamma \Lambda_n^2 + K_r \Lambda_n^h \Lambda_n^k Q + \eta^2 \T^4 n^4 \pi^4 \\ &  + \Lambda_n \T^2 n^2 \pi^2 \Bigl[ \gamma^2 (\eta+1)^2 +\gamma \Bigl( \frac{1}{k} \eta^2 +K_r \frac{1}{h} \Bigr) -\eta^2 J^2 \sigma_1^2 \Bigl] + \Bigl[ \eta^2 \Bigl( \frac{1}{k} +\gamma \Bigr) \Lambda_n^k + K_r \Gamma \Lambda_n^h \Bigr] \T^2 n^2 \pi^2 \\ & + J i \sigma_1 \Bigl\{ \Lambda_n^2 Q \gamma + \Lambda_n \Bigl[ Q K_r \Lambda_n^h + \gamma \Psi \Gamma + \T^2 n^2 \pi^2 \eta^2 \Bigl( 2 \gamma + \frac{1}{k} \Bigr) \Bigr] + K_r \Lambda_n^h \Lambda_n^k \Gamma + \eta^2 \Lambda_n^k \T^2 n^2 \pi^2 \Bigr\}, \\ g =& \Lambda_n (-J^2 \sigma_1^2 \Gamma +4 \gamma Q )+ Q \Psi + \T^2 n^2 \pi^2 \Bigl[ \gamma(\eta-1)^2+K_r \dfrac{1}{h} + \eta^2 \dfrac{1}{k} \Bigr] + J i \sigma_1 [ \Lambda_n (Q+4 \gamma \Gamma) + \Psi \Gamma + \eta^2 \T^2 n^2 \pi^2 ]. \end{aligned}$$
Hence, setting 
\begin{equation}\label{a,b} \begin{aligned} a_1 =& \Lambda_n \gamma \Psi Q + K_r \Lambda_n^h \Lambda_n^k Q + \eta^2 \T^4 n^4 \pi^4 + \Lambda_n \T^2 n^2 \pi^2 \Bigl[ \gamma^2 (\eta+1)^2 +\gamma \Bigl( \frac{1}{k} \eta^2 +K_r \frac{1}{h} \Bigr) \Bigl] \\ & + \Bigl[ \eta^2 \Bigl( \frac{1}{k} +\gamma \Bigr) \Lambda_n^k + K_r \Gamma \Lambda_n^h \Bigr] \T^2 n^2 \pi^2, \\ a_2 =& \Gamma \gamma \Lambda_n^2 + \Lambda_n \Lambda_n^h \Gamma K_r + \Lambda_n \T^2 n^2 \pi^2 \eta^2, \\ a_3 =& \Lambda_n^2 Q \gamma + \Lambda_n \Bigl[ Q K_r \Lambda_n^h + \gamma \Psi \Gamma + \T^2 n^2 \pi^2 \eta^2 \Bigl( 2 \gamma + \frac{1}{k} \Bigr) \Bigr] + K_r \Lambda_n^h \Lambda_n^k \Gamma + \eta^2 \Lambda_n^k \T^2 n^2 \pi^2, \\ b_1 =& 4 \gamma Q \Lambda_n +Q \Psi + \T^2 n^2 \pi^2 \Bigl[ \gamma(\eta-1)^2+K_r \dfrac{1}{h} + \eta^2 \dfrac{1}{k} \Bigr], \\ b_2 =& \Lambda_n \Gamma, \\ b_3 =& \Lambda_n (Q+4 \gamma \Gamma) + \Psi \Gamma + \eta^2 \T^2 n^2 \pi^2,
 \end{aligned} 
\end{equation}
from $(\ref{soglia2})$ one gets 
\begin{equation}
R = \dfrac{\Lambda_n+i \sigma_1}{a^2} \dfrac{a_1-J^2 \sigma_1^2 a_2 +i \sigma_1 J a_3}{b_1-J^2 \sigma_1^2 b_2 +i \sigma_1 J b_3},
\end{equation}
and consequently
\begin{equation}\label{soglia3}
R=\dfrac{\Lambda_n (\bar{a}+J^2 \sigma_1^2 a_3 b_3) - \sigma_1^2 J \bar{b} + i \sigma_1 (\bar{a} +J^2 \sigma_1^2 a_3 b_3 + \Lambda_n J \bar{b})}{a^2 [(b_1-J^2 \sigma_1^2 b_2)^2 +\sigma_1^2 J^2 b_3^2]},
\end{equation}
where $\bar{a}=(a_1-J^2 \sigma_1^2 a_2)(b_1-J^2 \sigma_1^2 b_2)$ and $\bar{b}=a_3(b_1-J^2 \sigma_1^2 b_2) -b_3(a_1-J^2 \sigma_1^2 a_2)$. Imposing the annulment of the imaginary part of $(\ref{soglia3})$, i.e. 
\begin{equation}
\bar{a} +J^2 \sigma_1^2 a_3 b_3 + \Lambda_n J \bar{b} =0,
\end{equation}
that is
\begin{equation}\label{sigma}
J^4 a_2 b_2 \sigma_1^4 -J^2 \sigma_1^2 [a_2 b_1 + a_1 b_2 - a_3 b_3 + \Lambda_n J (a_3 b_2 - a_2 b_3)] + a_1 b_1 + \Lambda_n J (a_3 b_1 - b_3 a_1) =0,
\end{equation}
the critical Rayleigh number for the onset of the oscillatory convection is given by:  
\begin{equation}\label{osc}
\!\! R_O \!=\! \min_{(n,a^2) \in \N \times \R^+} \! \dfrac{\Lambda_n [ (a_1 \!-\! J^2 \sigma_1^2 a_2)(b_1 \!-\! J^2 \sigma_1^2 b_2)+J^2 \sigma_1^2 a_3 b_3 ] \!-\! \sigma_1^2 J [a_3(b_1 \!-\! J^2 \sigma_1^2 b_2) \!-\! b_3(a_1 \!-\! J^2 \sigma_1^2 a_2)] }{a^2 [(b_1-J^2 \sigma_1^2 b_2)^2 +\sigma_1^2 J^2 b_3^2]},
\end{equation}
where $\sigma_1^2$ is the positive root of $(\ref{sigma})$. Let us point out that if 
\begin{equation}
[a_2 b_1 + a_1 b_2 - a_3 b_3 + \Lambda_n J (a_3 b_2 - a_2 b_3)]^2 - 4 a_2 b_2 [a_1 b_1 + \Lambda_n J (a_3 b_1 - b_3 a_1)]<0,
\end{equation}
or if 
\begin{equation}
a_2 b_1 + a_1 b_2 - a_3 b_3 + \Lambda_n J (a_3 b_2 - a_2 b_3)<0, \quad a_1 b_1 + \Lambda_n J (a_3 b_1 - b_3 a_1)>0,
\end{equation}
oscillatory convection cannot occur. Due to its complicated algebraic form, the minimization $(\ref{osc})$ will be numerically investigated in section \ref{numerical}.  \\ \\
Neglecting the Vadasz term ($J=0$), from $(\ref{osc})$ one gets 
\begin{equation}
R_O=\min_{n,a^2} \dfrac{\Lambda_n a_1}{a^2 b_1},
\end{equation}
and hence $R_O=R_S$, which coincides with the result found in \cite{legg}. 


\section{Numerical simulations}\label{numerical}
\begin{figure}[h!]
\centering
\subfigure[$h>>k$]{
\includegraphics[scale=0.35]{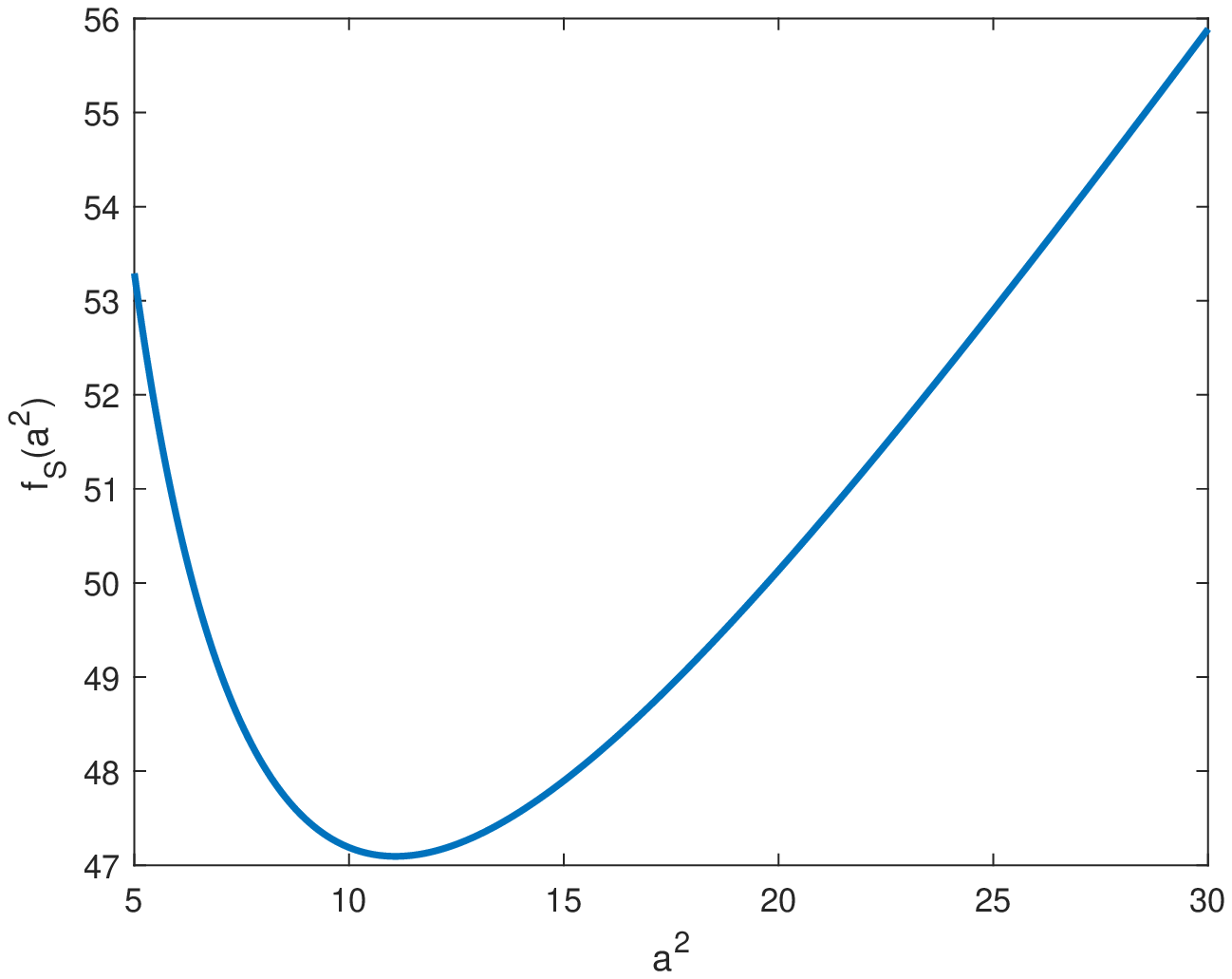}}
\subfigure[$h=k$]{
\includegraphics[scale=0.35]{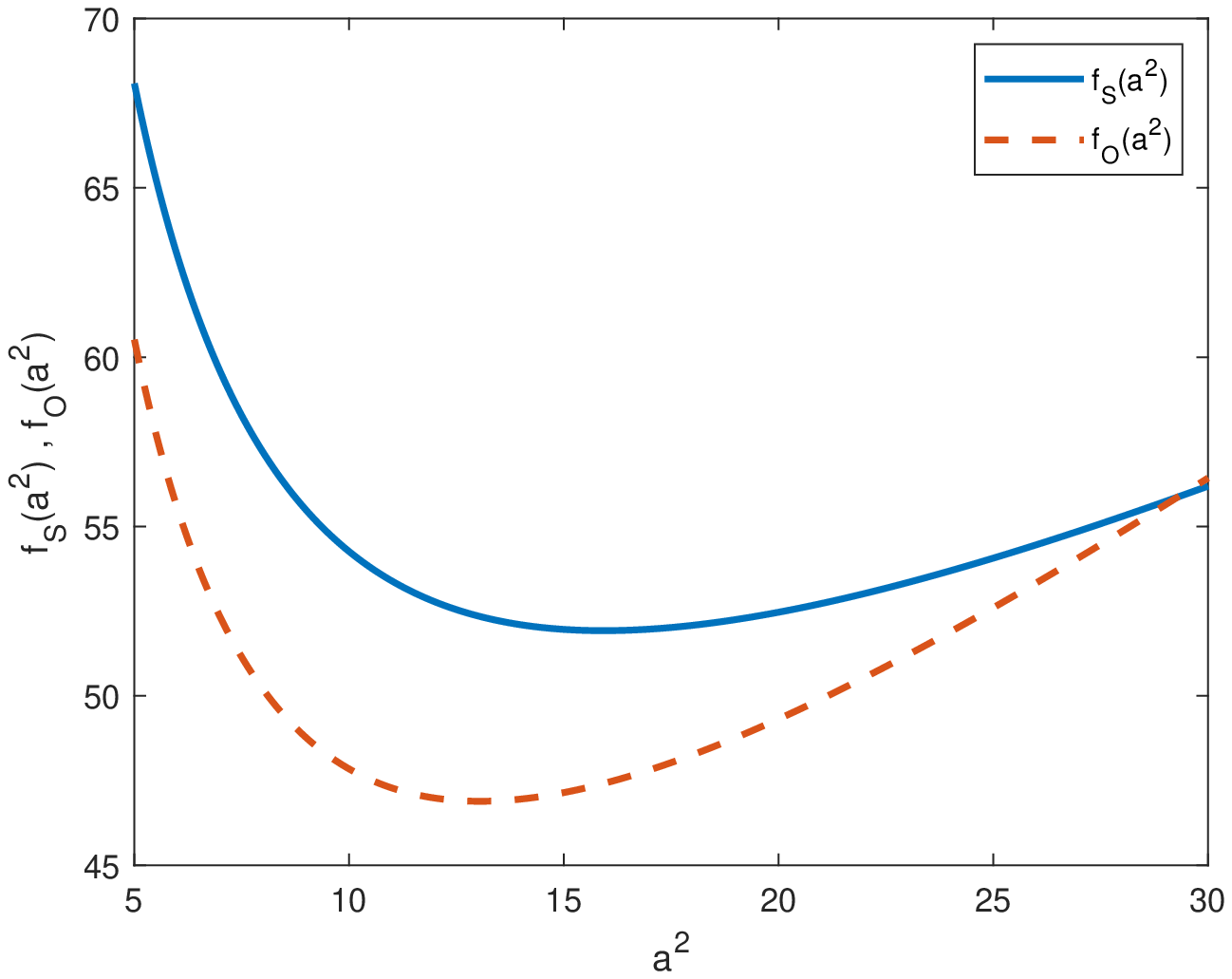}}
\subfigure[$h<<k$]{
\includegraphics[scale=0.35]{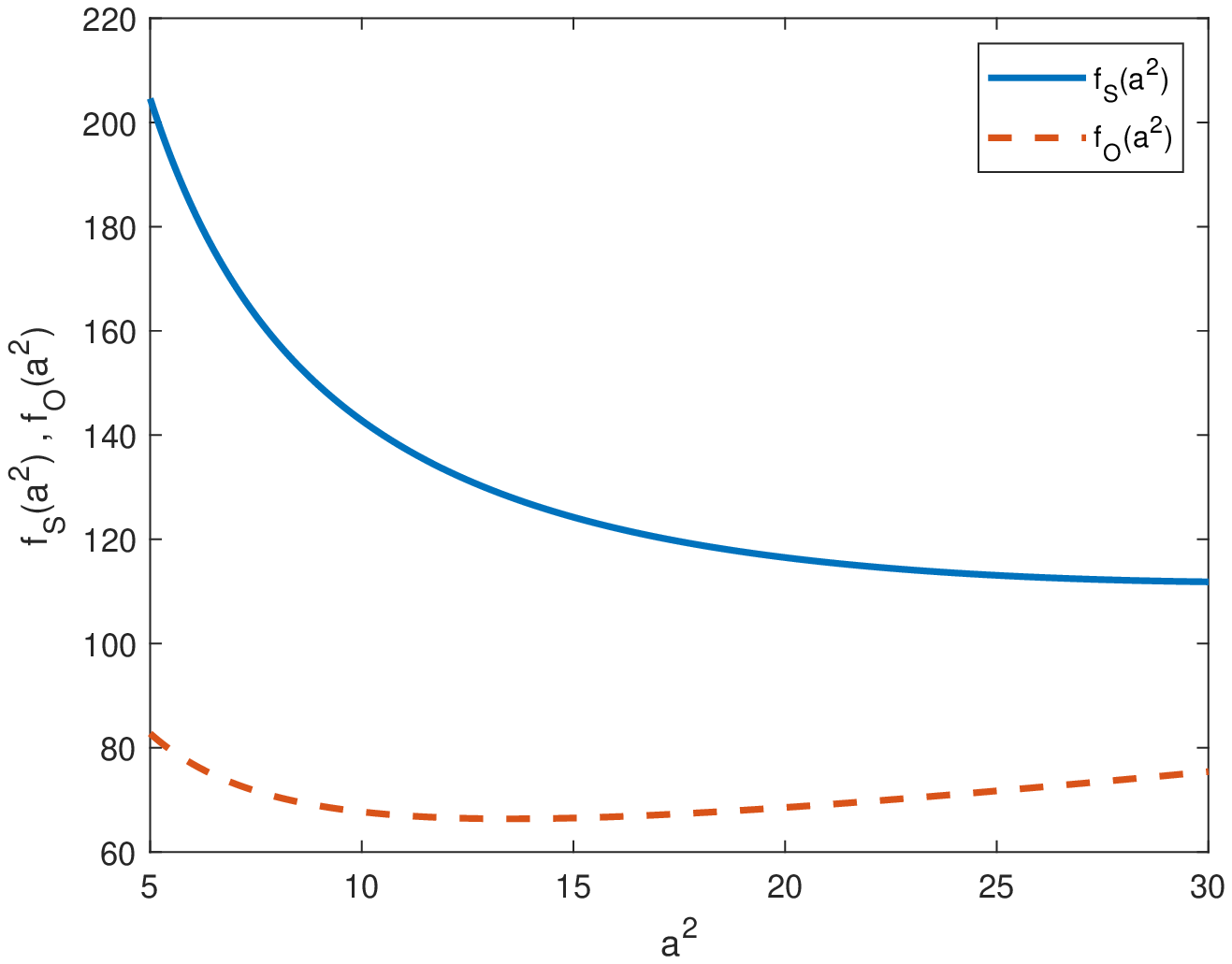}}
\caption{Instability thresholds for quoted values of $k$ and $h$ and for $\T^2=10,{\bf J=0.5},K_r=1.5,\eta=0.2,\gamma=0.8$. (a): $h=10,k=0.1$, (b): $h=k=1$, (c): $h=0.1,k=10$.}
\label{FIGURA1}
\end{figure}

\begin{figure}[!h]
\centering
\subfigure[$h>>k$]{
\includegraphics[scale=0.35]{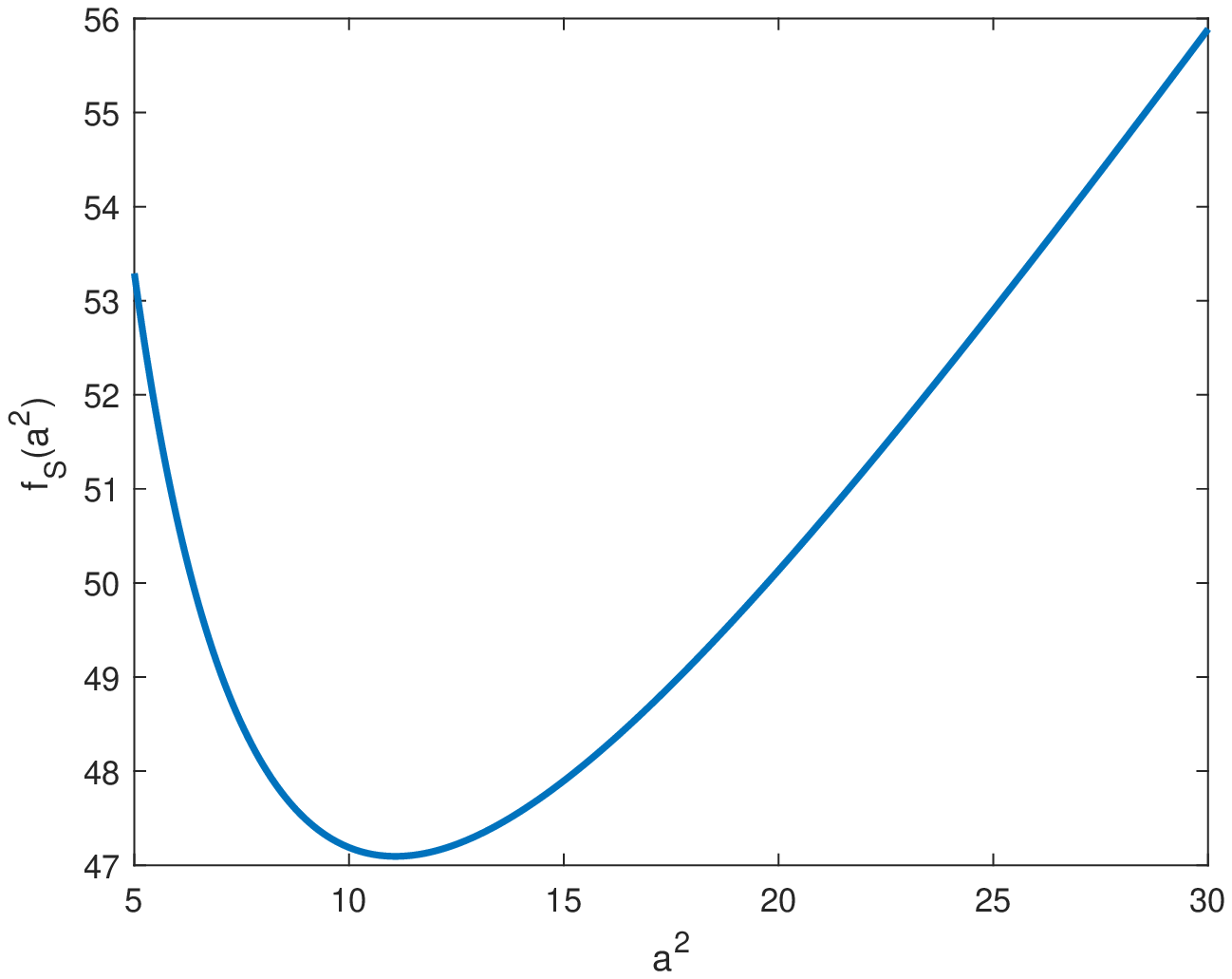}}
\subfigure[$h=k$]{
\includegraphics[scale=0.35]{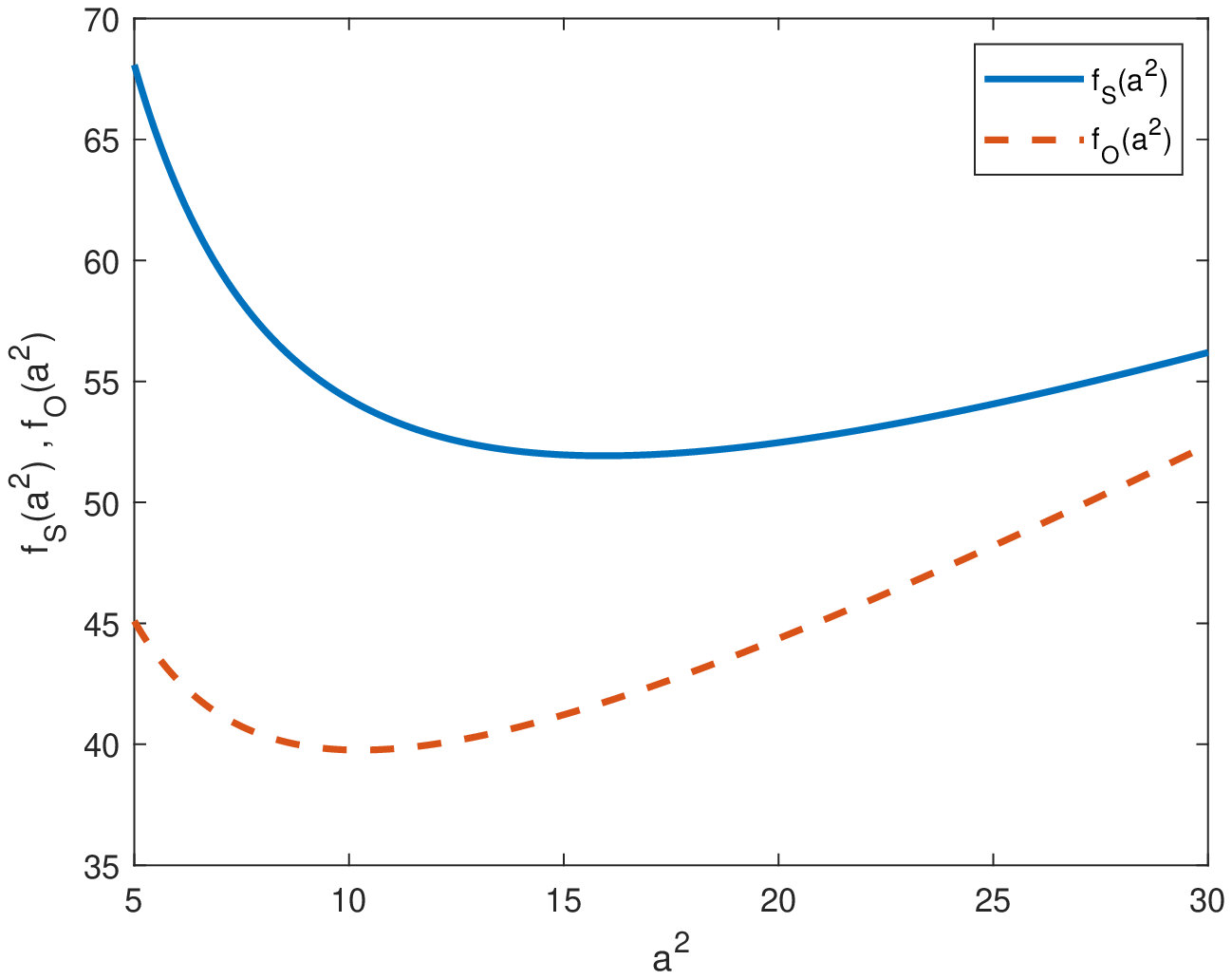}}
\subfigure[$h<<k$]{
\includegraphics[scale=0.35]{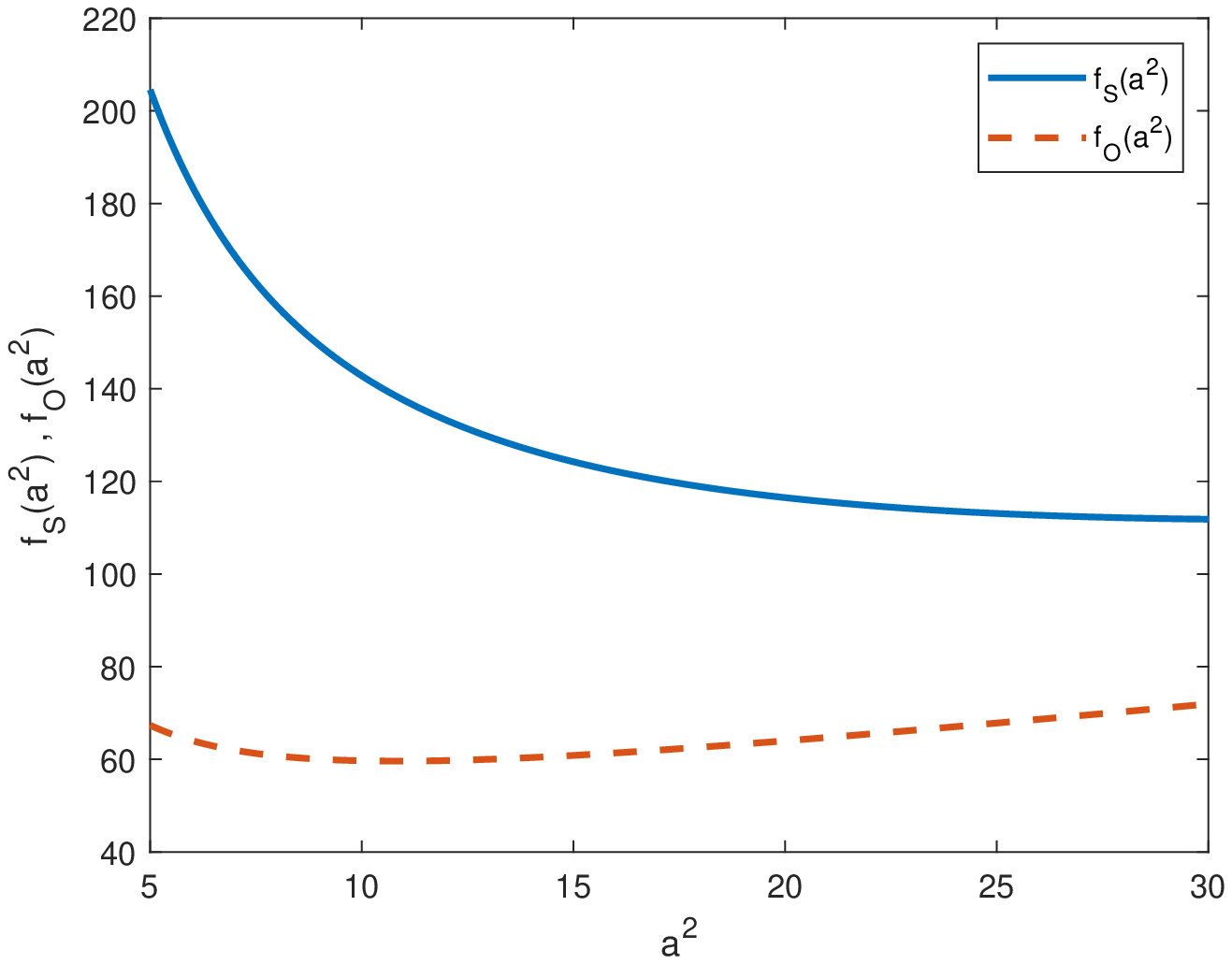}}
\caption{Instability thresholds for quoted values of $k$ and $h$ and for $\T^2=10,{\bf J=10},K_r=1.5,\eta=0.2,\gamma=0.8$. (a): $h=10,k=0.1$, (b): $h=k=1$, (c): $h=0.1,k=10$.}
\label{FIGURA2}
\end{figure}

\begin{figure}[h!]
\centering
\subfigure[]{
\includegraphics[scale=0.5]{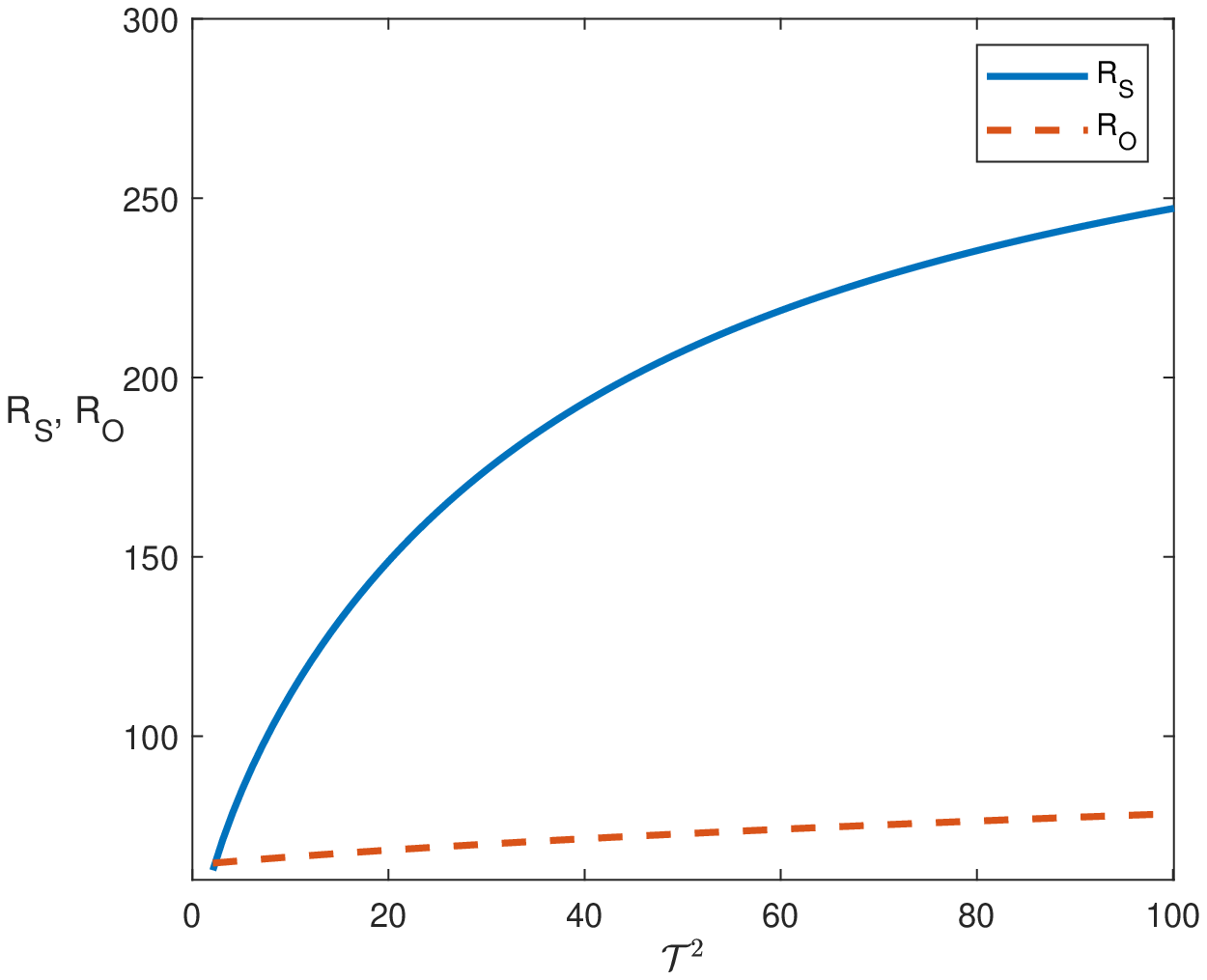}}
\subfigure[]{
\includegraphics[scale=0.5]{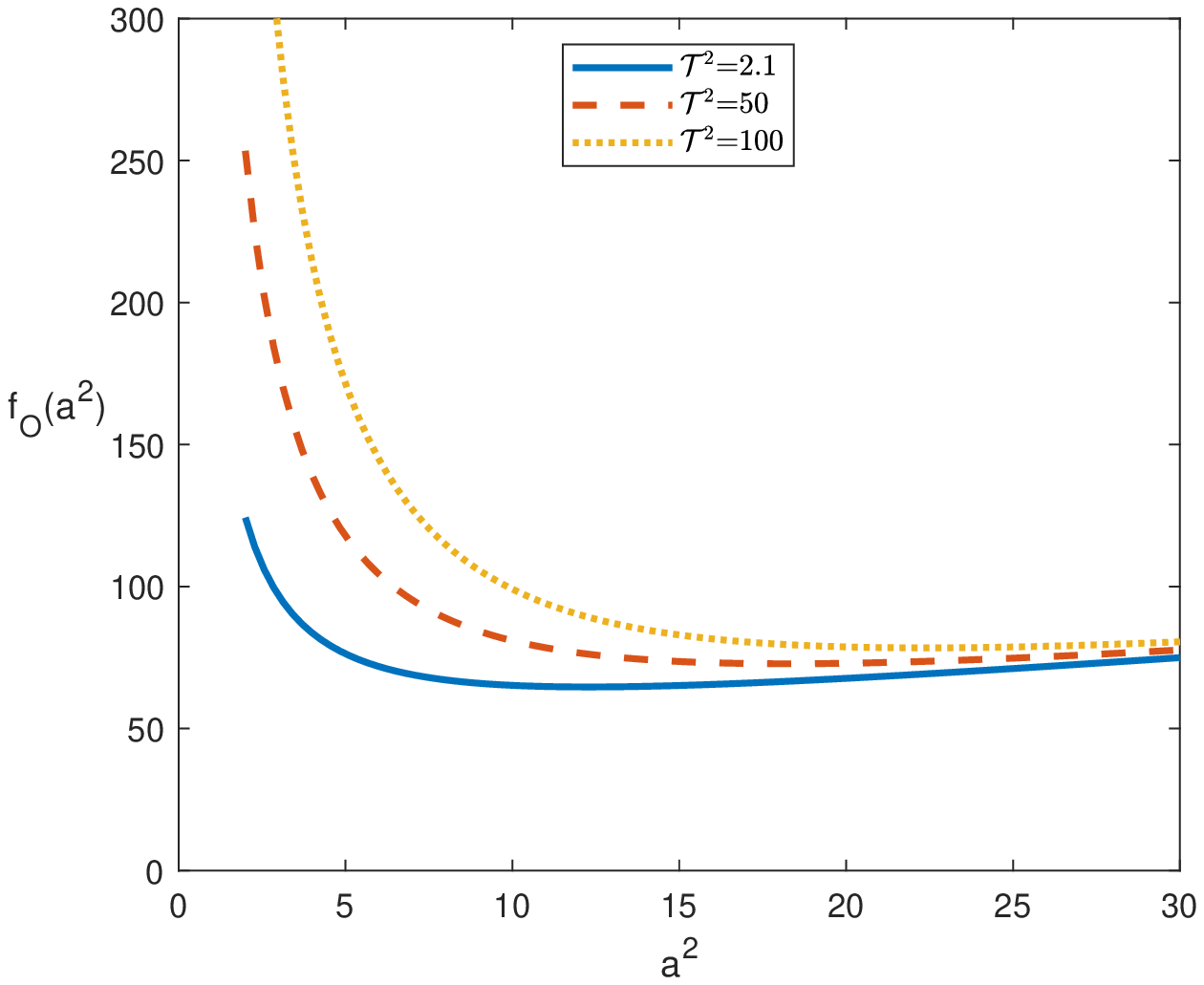}}
\caption{(a): Stationary and Oscillatory instability thresholds as functions of the Taylor number $\T^2$ for $k=10,h=0.1,J=0.5,K_r=1.5,\eta=0.2,\gamma=0.8$. (b): Oscillatory thresholds at quoted values of the Taylor number $\T^2$ and for $k=10,h=0.1,J=0.5,K_r=1.5,\eta=0.2,\gamma=0.8$.}
\label{FIGURA3}
\end{figure}

\begin{figure}[h!]
\centering
\subfigure[]{
\includegraphics[scale=0.5]{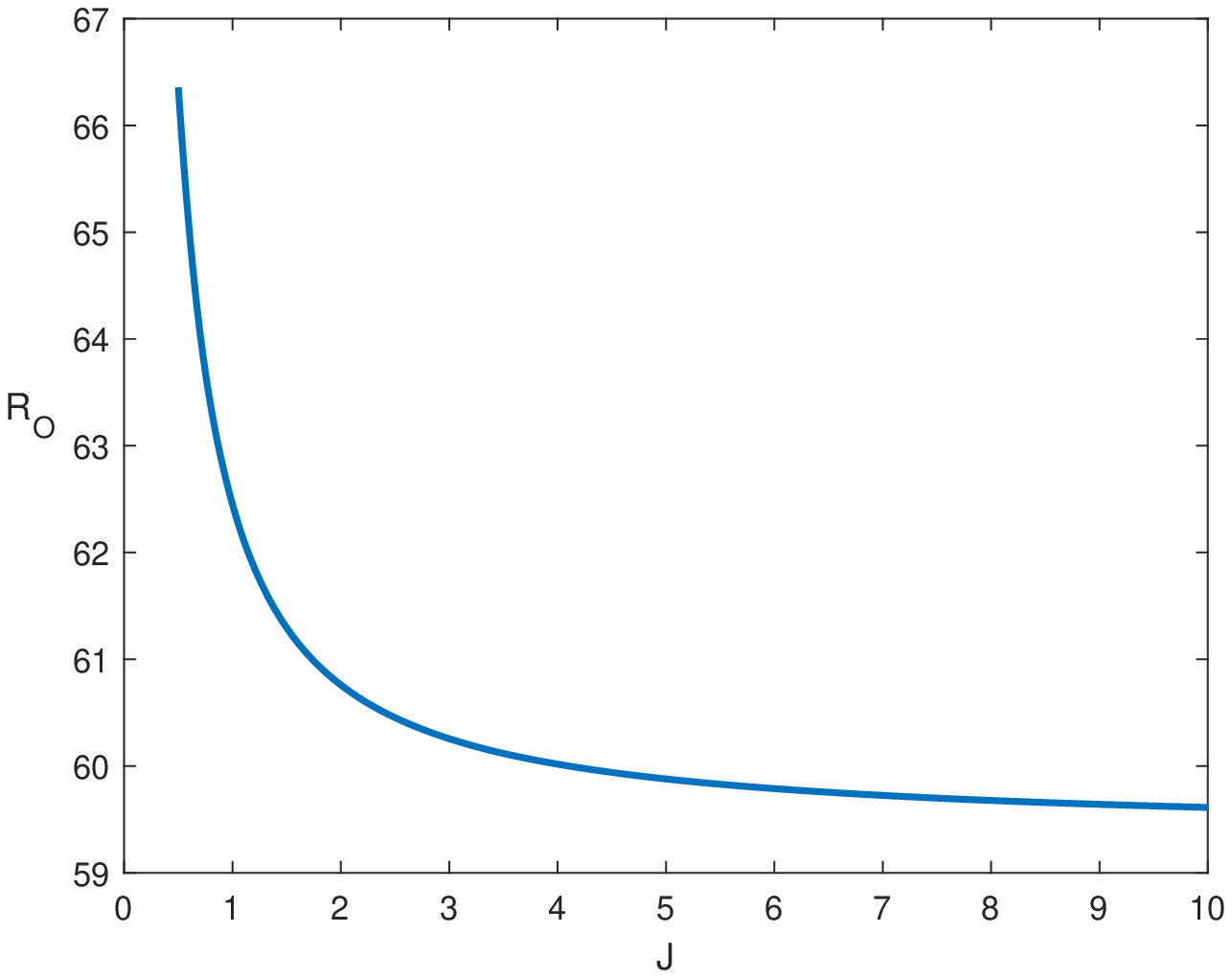}}
\subfigure[]{
\includegraphics[scale=0.5]{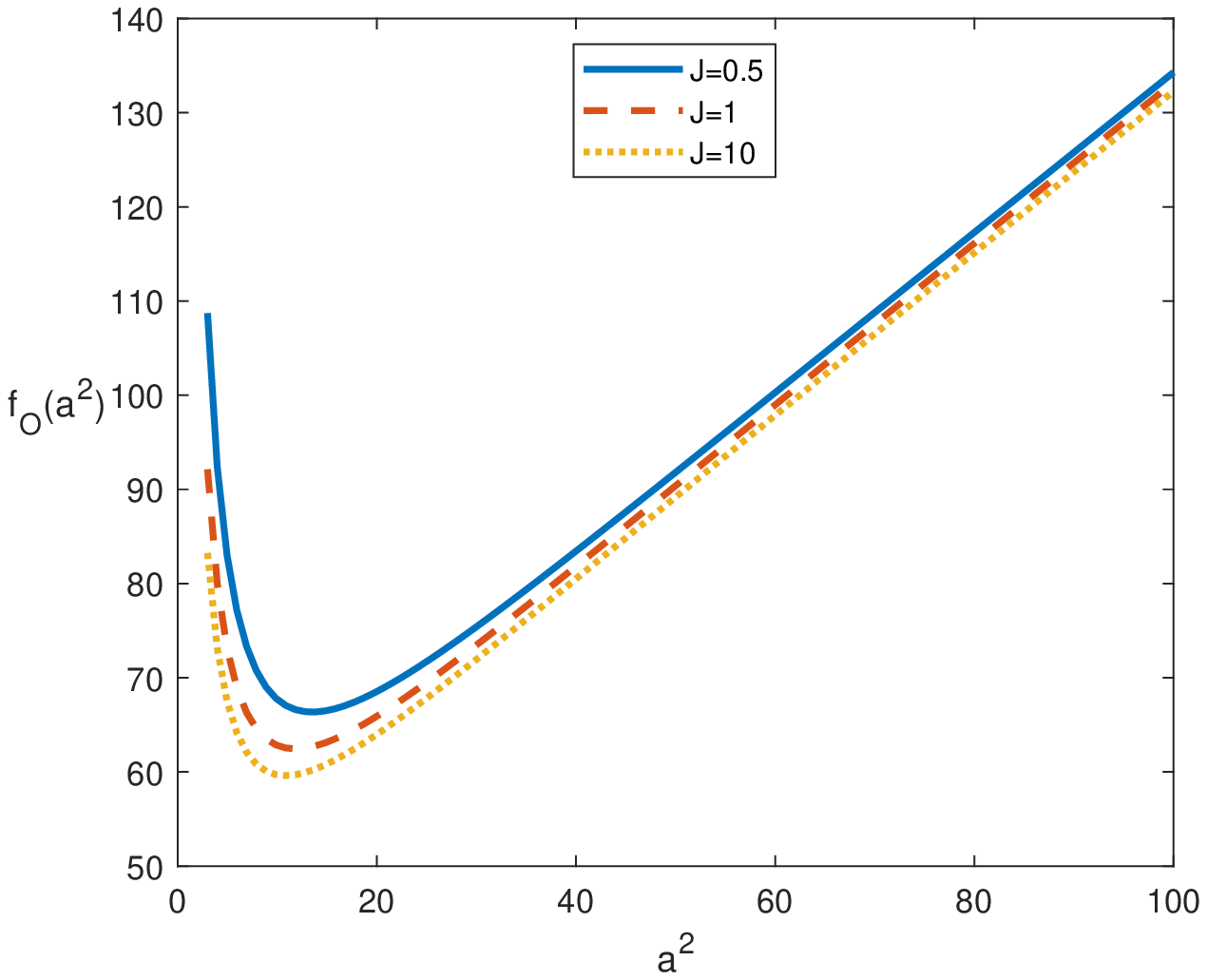}}
\caption{(a): Oscillatory instability threshold as function of the Vadasz term $J$ for $k=10,h=0.1,\T^2=10,K_r=1.5,\eta=0.2,\gamma=0.8$. (b): Oscillatory thresholds at quoted values of the Vadasz term $J$ and for $k=10,h=0.1,\T^2=10,K_r=1.5,\eta=0.2,\gamma=0.8$.}
\label{FIGURA4}
\end{figure}

\begin{figure}[h!]
\centering
\subfigure[]{
\includegraphics[scale=0.5]{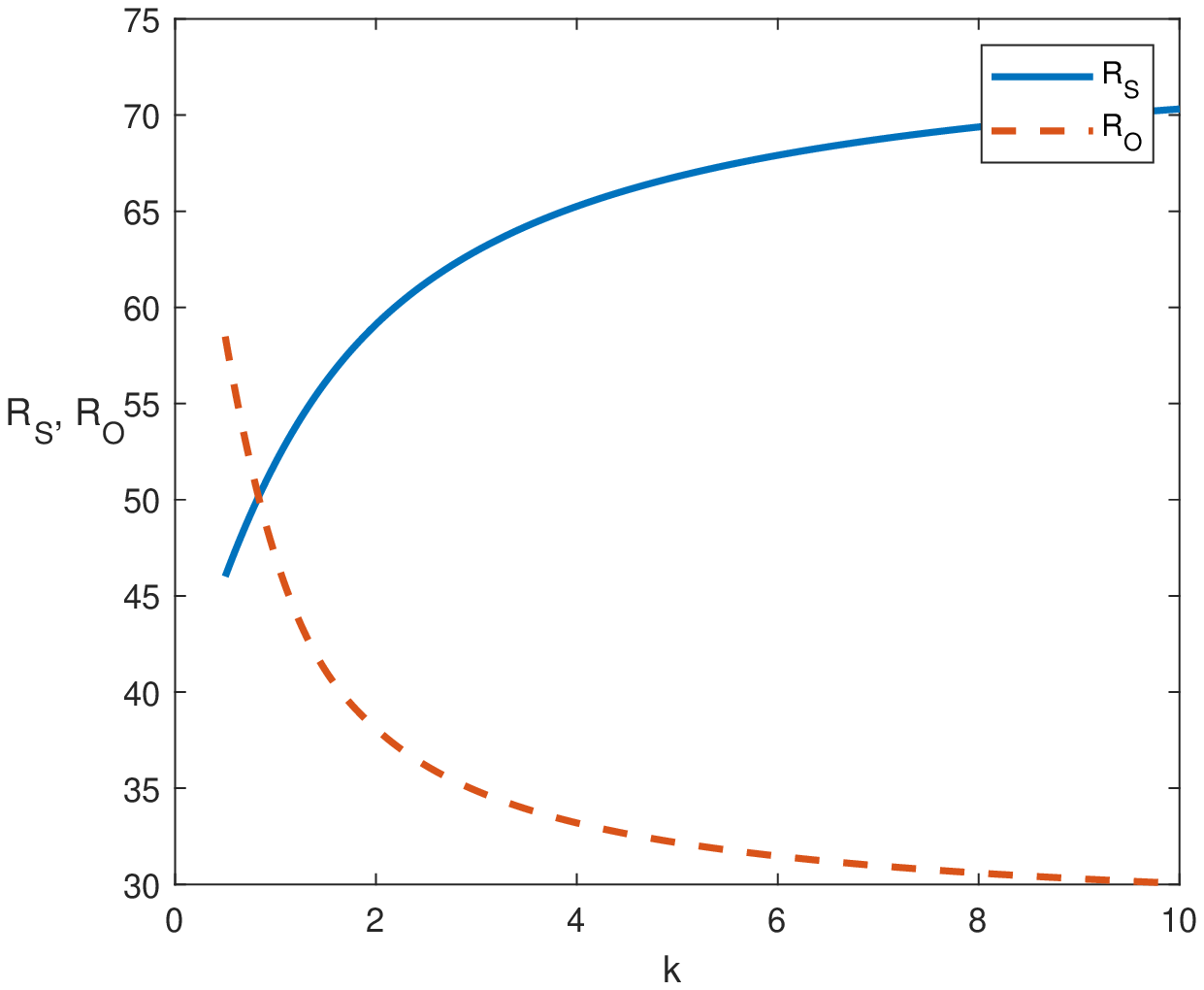}}
\subfigure[]{
\includegraphics[scale=0.5]{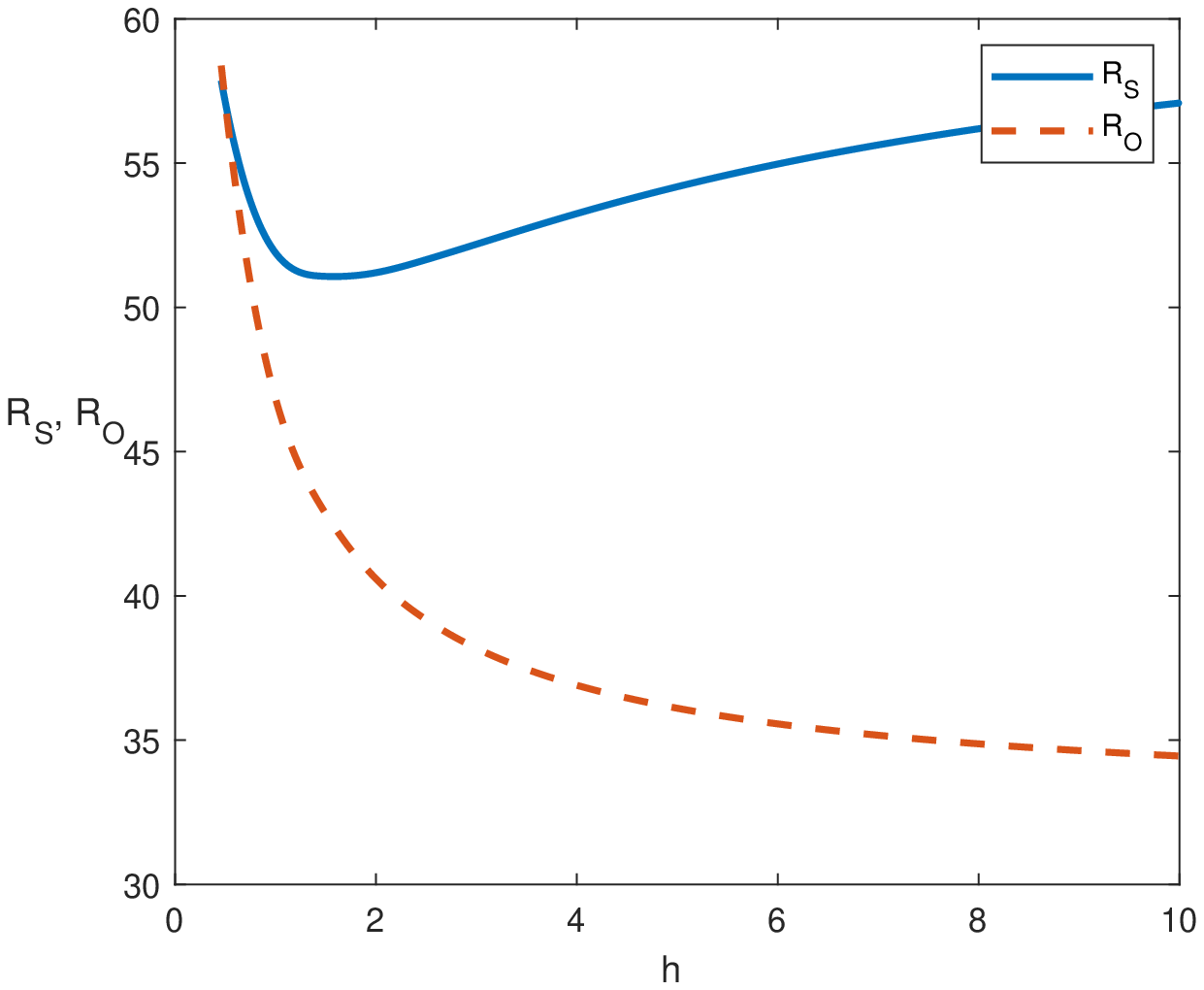}}
\caption{Asymptotic behaviour of the steady and oscillatory instability thresholds as functions of the anisotropic permeability parameters $k$ and $h$. (a): $h=1,\T^2=10,J=0.5,K_r=1.5,\eta=0.2,\gamma=0.8$. (b): $k=1,\T^2=10,J=0.5,K_r=1.5,\eta=0.2,\gamma=0.8$.}
\label{FIGURA5}
\end{figure}

The aim of this section is to solve $(\ref{staz})$ and $(\ref{osc})$ and numerically describe the asymptotic behaviour of the steady and oscillatory critical Rayleigh numbers with respect to $\T^2$, $J$, $k$, $h$, in order to describe the influence of rotation, Vadasz term, anisotropic macropermeability and anisotropic micropermeability on the onset of convection. Through numerical simulations, we have shown that the minimum of both $(\ref{staz})$ and $(\ref{osc})$ with respect to $n$ is attained at $n=1$, hence let us define
\begin{equation}\label{fs}
f_S(a^2) = \dfrac{\Lambda_1}{a^2} \dfrac{\gamma ( \Lambda_1^k \!+\! K_r \Lambda_1^h ) H_S \Lambda_1 \!+\! K_r H_S \Lambda_1^k \Lambda_1^h \!+\! M_S \Lambda_1 \T^2 \pi^2 \!+\! \eta^2 \T^4 \pi^4 \!+\! N_S \T^2 \pi^2}{H_S 4 \gamma \Lambda_1+H_S (\Lambda_1^k+K_r\Lambda_1^h)+E_S \T^2 \pi^2} \ ,
\end{equation}
and
\begin{equation}\label{fo}
f_O(a^2) = \dfrac{\Lambda_1 (\bar{a}+J^2 \sigma_1^2 a_3 b_3) - \sigma_1^2 J \bar{b} }{a^2 [(b_1-J^2 \sigma_1^2 b_2)^2 +\sigma_1^2 J^2 b_3^2]} \ ,
\end{equation}
where $a_i$ and $b_i$, for $i=1,2,3$, are now given by $(\ref{a,b})$ for $n=1$.
In all numerical simulations we have set $\{K_r=1.5,\gamma=0.8,\eta=0.2\}$ \cite{inertia}. Varying $k,h,\T^2,J$ in turn we have found some critical values of these parameters for which oscillatory convection cannot occur. In particular
\begin{itemize}
\item from table $\ref{TT1}(a)$, for large values of the Vadasz term ($J=10$), we find out that exists $k^* \in (0.53,0.54)$ such that if $k<k^*$ convection can arise only through stationary motion; for $k \in (0.57,0.58)$ there is a switch from steady to oscillatory convection; 
\item table $\ref{TT1}(b)$ shows that for large values of $J$ and for increasing $h$, if convection occurs, it can set in only via an oscillatory state;  
\item table $\ref{TT2}(a)$ displays that there is a similar behaviour between the small Vadasz term case $(J=0.5)$ and the large $J$ case, but for smaller $J$ the threshold $k^{**}$ for which oscillatory convection cannot occur is higher, in particular there exists $k^{**} \in (0.82,0.83)$ such that if $k<k^{**}$ convection can arise only via stationary motion, while for $k \in (0.83,0.84)$ there is a transition from steady to oscillatory convection; 
\item for small Vadasz term $J$, from table $\ref{TT2}(b)$ we find a threshold $h^* \in (0.45,0.46)$ such that if $h<h^*$ oscillatory convection cannot occur; for $h \in (0.49,0.5)$ there is a reversal from steady to oscillatory convection;  
\item from tables $\ref{TT1}$ and $\ref{TT2}$ it arises that $R_O$ decreases for both increasing $k$ and $h$, hence, when convection occurs via oscillatory motion, the system become more unstable at increasing permeability parameters; 
\item tables $\ref{TT3}$ display a peculiar situation: for $h<<k$ convection sets in only via oscillatory motion, except for very small Vadasz term $J<0.11$, for which stationary convection occurs; for $h>>k$ oscillatory convection cannot arise (in particular, for $\{h=0.1,k=10,\T^2=10\}$ the steady critical Rayleigh number is $R_S=111.7045$, while for $\{h=10,k=0.1,\T^2=10\}$ the steady critical Rayleigh number is $R_S=47.0944$);  
\item from table $\ref{TT3}(a)$ we may remark that for $h<<k$ $R_O$ is a decreasing function of $J$, as already observed in \cite{inertia} for $h=k=1$; 
\item from table $\ref{TT4}(a)$ we notice that if $h<<k$ there exists a threshold $\T^{2*} \in (2.09,2.1)$ such that for $\T^2<\T^{2*}$ oscillatory convection cannot arise and stationary convection sets in for $\T^2$ up to $2.3$, while for $\T^2 \in (2.3,2.4)$ there is a switch from stationary to oscillatory convection; from table $\ref{TT4}(b)$ we recover the same behaviour shown in table $\ref{TT3}(b)$;
\item tables $\ref{TT4}$ numerically show off the stabilizing effect of rotation on the onset of convection, since both $R_S$ and $R_O$ are increasing functions with respect to $\T^2$, as one is expected; in particular $R_O$, if it exists, has a \textit{slower} increase with respect to $\T^2$ than $R_S$.   
\end{itemize} 
In figures $\ref{FIGURA1}$ and $\ref{FIGURA2}$ the instability thresholds at quoted values of the permeability parameters $h$ and $k$ are shown, for small and large Vadasz term $J$, respectively. From figure $\ref{FIGURA3}$ we may visualize the stabilizing effect of the Taylor number $\T^2$ on the onset of convection, in particular from figure $\ref{FIGURA3}(a)$ two very different growth rates of the steady and of the oscillatory instability thresholds arise. Figure $\ref{FIGURA4}$ shows the destabilizing effect on the onset of oscillatory convection of the Vadasz coefficient $J$. The numerical results of table $\ref{TT2}$ are graphically shown in figure $\ref{FIGURA5}$, where steady and oscillatory instability thresholds are represented as functions of the anisotropic permeability parameters $k$ and $h$.

\begin{table}[h]
\centering{
\subtable[]{\label{TT1a}
\begin{tabular}{|c|c|c|c|c|}
\hline
$R_S$ & $R_O$ & $a^2_S$ & $a^2_O$ & $k$ \\
\hline
55.5802 & $\not\exists$ & 15.0097 & $\not\exists$ & 0.1 \\
46.4633 & $\not\exists$ & 15.4106 & $\not\exists$ & 0.5 \\
46.7672 & $\not\exists$ & 15.4530 & $\not\exists$ & 0.53 \\
46.8724 & 48.0833 & 15.4670 & 10.8010 & 0.54 \\
46.9790 & 47.7998 & 15.4809 & 10.7872 & 0.55 \\
47.0871 & 47.5233 & 15.4947 & 10.7733 & 0.56 \\
47.1964 & 47.2537 & 15.5084 & 10.7596 & 0.57 \\
47.3067 & 46.9906 & 15.5220 & 10.7459 & 0.58 \\
47.5300 & 46.4833 & 15.5487 & 10.7186 & 0.6 \\
51.9256 & 39.7591 & 15.9190 & 10.2433 & 1 \\
66.8036 & 28.7339 & 15.6424 & 8.7834 & 5 \\
70.3183 & 27.0015 & 15.3278 & 8.4389 & 10 \\
\hline
\end{tabular}
\qquad
}
\subtable[]{\label{TT1b}
\begin{tabular}{|c|c|c|c|c|}
\hline
$R_S$ & $R_O$ & $a^2_S$ & $a^2_O$ & $h$ \\
\hline
146.5652 & 133.6558 & 22.0926 & 10.6720 & 0.01 \\
88.7111 & 86.2352 & 26.9067 & 14.1899 & 0.1 \\
56.9286 & 48.5663 & 18.8637 & 11.9393 & 0.5 \\
51.9256 & 39.7591 & 15.9190 & 10.2433 & 1 \\
54.1833 & 31.1656 & 13.6497 & 8.3863 & 5 \\
57.0855 & 30.0809 & 13.6502 & 8.2253 & 10 \\
61.7553 & 29.1760 & 13.8544 & 8.1706 & 100 \\
\hline
\end{tabular}
} }
\caption{Critical steady and oscillatory Rayleigh numbers and wavenumbers for quoted values of $k$ (a) and for quoted values of $h$ (b). \textit{Table a}: $h=1,\T^2=10,J=10,K_r=1.5,\eta=0.2,\gamma=0.8$. \textit{Table b}: $k=1,\T^2=10,J=10,K_r=1.5,\eta=0.2,\gamma=0.8$.}
\label{TT1}
\end{table}

\begin{table}[h]
\centering{
\subtable[]{\label{TT2a}
\begin{tabular}{|c|c|c|c|c|}
\hline
$R_S$ & $R_O$ & $a^2_S$ & $a^2_O$ & $k$ \\
\hline
55.5802 & $\not\exists$ & 15.0097 & $\not\exists$ & 0.1 \\
46.4633 & $\not\exists$ & 15.4106 & $\not\exists$ & 0.5 \\
49.8064 & $\not\exists$ & 15.7735 & $\not\exists$ & 0.8 \\
50.0286 & $\not\exists$ & 15.7914 & $\not\exists$ & 0.82 \\
50.1389 & 50.2184 & 15.8001 & 13.4683 & 0.83 \\
50.2488 & 49.9896 & 15.8085 & 13.4404 & 0.84 \\
50.3581 & 49.7655 & 15.8168 & 13.4130 & 0.85 \\
50.8960 & 48.7116 & 15.8553 & 13.2826 & 0.9 \\ 
51.9256 & 46.8876 & 15.9190 & 13.0503 &  1 \\
66.8036 & 32.1661 & 15.6424 & 10.6842 & 5 \\
70.3183 & 30.0517 & 15.3278 & 10.2302 & 10 \\
\hline
\end{tabular}
\qquad
}
\subtable[]{\label{TT2b}
\begin{tabular}{|c|c|c|c|c|}
\hline
$R_S$ & $R_O$ & $a^2_S$ & $a^2_O$ & $h$ \\
\hline
88.7111 & $\not\exists$ & 26.9067 & $\not\exists$ & 0.1 \\
59.6059 & $\not\exists$ & 20.0551 & $\not\exists$ & 0.4 \\
58.1248 & $\not\exists$ & 19.4148 & $\not\exists$ & 0.45 \\
57.8653 & 58.3814 & 19.2980 & 15.3814 & 0.46 \\
57.6164 & 57.9856 & 19.1847 & 15.3599 & 0.47 \\
57.3777 & 57.6028 & 19.0746 & 15.2905 & 0.48 \\
57.1486 & 57.2324 & 18.9676 & 15.2227 & 0.49 \\
56.9286 & 56.8737 & 18.8637 & 15.1563 & 0.5 \\
51.9256 & 46.8876 & 15.9190 & 13.0503 &  1 \\
54.1833 & 36.1050 & 13.6497 & 10.4542 & 5 \\
57.0855 & 34.4490 & 13.6502 & 10.1144 & 10 \\
\hline
\end{tabular}
} }
\caption{Critical steady and oscillatory Rayleigh numbers and wavenumbers for quoted values of $k$ (a) and for quoted values of $h$ (b). \textit{Table a}: $h=1,\T^2=10,J=0.5,K_r=1.5,\eta=0.2,\gamma=0.8$. \textit{Table b}: $k=1,\T^2=10,J=0.5,K_r=1.5,\eta=0.2,\gamma=0.8$.}
\label{TT2}
\end{table}

\begin{table}[!h]
\centering{
\subtable[$h<<k$]{\label{TT3a}
\begin{tabular}{|c|c|c|}
\hline
$R_O$ & $a^2_O$ & $J$ \\
\hline
$\not\exists$ & $\not\exists$ & 0 \\
$\not\exists$ & $\not\exists$ & 0.1 \\
94.6887 & 27.8062 & 0.11 \\
85.6326 & 23.1078 & 0.15 \\
66.3568 & 13.5068 & 0.5 \\
62.4493 & 11.8203 & 1 \\
59.8788 & 10.9098 & 5 \\
59.6123 & 10.8360 & 10 \\
59.4095 & 10.7843 & 50 \\
59.3849 & 10.7784 & 100 \\
\hline
\end{tabular}
\qquad
}
\subtable[$h>>k$]{\label{TT3b}
\begin{tabular}{|c|c|c|}
\hline
$R_O$ & $a^2_O$ & $J$ \\
\hline
$\not\exists$ & $\not\exists$ & 0 \\
$\not\exists$ & $\not\exists$ & 0.5 \\
$\not\exists$ & $\not\exists$ & 1 \\
$\not\exists$ & $\not\exists$ & 5 \\
$\not\exists$ & $\not\exists$ & 10 \\
$\not\exists$ & $\not\exists$ & 50 \\
$\not\exists$ & $\not\exists$ & 100 \\
\hline
\end{tabular}
} }
\caption{Critical oscillatory Rayleigh numbers and wavenumbers for quoted values of $J$ for $h<<k$ (a) and for $h>>k$ (b). \textit{Table a}: for $h=0.1,k=10,\T^2=10,K_r=1.5,\eta=0.2,\gamma=0.8$ (for these values, critical steady Rayleigh number and wavenumber are $R_S=111.7045,a^2_S=32.4936$). \textit{Table b}: for $h=10,k=0.1,\T^2=10,K_r=1.5,\eta=0.2,\gamma=0.8$ (for these values, critical steady Rayleigh number and wavenumber are $R_S=47.0944,a^2_S=11.0770$).}
\label{TT3}
\end{table}

\begin{table}[h]
\centering{
\subtable[$h<<k$]{\label{TT4a}
\begin{tabular}{|c|c|c|c|c|}
\hline
$R_S$ & $R_O$ & $a^2_S$ & $a^2_O$ & $\T^2$ \\
\hline
37.9913 & $\not\exists$ & 8.8113 & $\not\exists$ & 0 \\
61.7538 & $\not\exists$ & 17.9673 & $\not\exists$ & 2 \\
62.5794 & $\not\exists$ & 18.2713 & $\not\exists$ & 2.09 \\
62.6704 & 64.6393 & 18.3047 & 12.2819 & 2.1 \\
63.5719 & 64.6626 & 18.6347 & 12.2984 & 2.2 \\
64.4590 & 64.6858 & 18.9577 & 12.3149 & 2.3 \\
65.3325 & 64.7090 & 19.2739 & 12.3313 & 2.4 \\
66.1929 & 64.7321 & 19.5835 & 12.3477 & 2.5 \\
84.6085 & 65.2968 & 25.6964 & 12.7489 & 5 \\
111.7045 & 66.3568 & 32.4936 & 13.5068 & 10 \\
\hline
\end{tabular}
\qquad
}
\subtable[$h>>k$]{\label{TT4b}
\begin{tabular}{|c|c|c|c|c|}
\hline
$R_S$ & $R_O$ & $a^2_S$ & $a^2_O$ & $\T^2$ \\
\hline
39.6844 & $\not\exists$ & 8.8024 & $\not\exists$ & 0 \\
43.4949 & $\not\exists$ & 10.0015 & $\not\exists$ & 5 \\
47.0944 & $\not\exists$ & 11.0770 & $\not\exists$ & 10 \\
100.1927 & $\not\exists$ & 22.9601 & $\not\exists$ & 100 \\
\hline
\end{tabular}
} }
\caption{Critical steady and oscillatory Rayleigh numbers and wavenumbers for quoted values of $\T^2$ for $h<<k$ (a) and for $h>>k$ (b). \textit{Table a}: for $h=0.1,k=10,J=0.5,K_r=1.5,\eta=0.2,\gamma=0.8$. \textit{Table b}: for $h=10,k=0.1,J=0.5,K_r=1.5,\eta=0.2,\gamma=0.8$.}
\label{TT4}
\end{table}

\section{Conclusions}
The onset of convection in a rotating and anisotropic BDPM, taking into account the Vadasz term, has been studied via linear instability analysis. Let us remark that the Vadasz term allows the onset of oscillatory convection, which is not present when the inertia is neglected (see \cite{legg}). Moreover, if $h=k=1$, i.e. confining ourselves to the isotropic case, from $(\ref{staz})$ and $(\ref{osc})$ we recover the stationary and oscillatory thresholds found in \cite{inertia}, respectively. Lastly, it has been numerically investigated the relationship between the critical steady and oscillatory Rayleigh numbers and the fundamental parameters $h,k,\T^2,J$ and we have found out that: 
\begin{itemize}
\item $R_S$ and $R_O$ are increasing functions of $\T^2$; 
\item $R_O$, if it exists, is a decreasing function of $J$;
\item comparing our results with those ones found in \cite{inertia}, anisotropic macropermeability and anisotropic micropermeability lead to higher steady and oscillatory thresholds.
\end{itemize} 

\textbf{Acknowledgements.} This paper has been performed under the auspices of the GNFM of INdAM.

\end{document}